\documentclass[twocolumn]{aastex701} 
\definecolor{trackchange}{cmyk}{0.5,0,0,0.5}

\usepackage[normalem]{ulem} 

\usepackage{graphics,epsf}
\usepackage[utf8]{inputenc}
\usepackage{amsmath}                
\usepackage{amsfonts}               
\usepackage{amssymb}                
\usepackage{epsfig}                 
\usepackage{graphicx}               
\usepackage{float}
\usepackage{color}
\usepackage{multirow}               
\usepackage{hyperref}

\hypersetup{
    colorlinks=true,
    linkcolor=red,   
    urlcolor=cyan}

\usepackage[colorinlistoftodos]{todonotes}
\usepackage{todonotes}

\newcommand{\kms}{{~\rm km\; s^{-1}}}

\newcommand{\km}{{~\rm km}}
\newcommand{\s}{{~\rm s}}

\newcommand{\g}{{~\rm g}}
\newcommand{\G}{{~\rm G}}
\newcommand{\K}{{~\rm K}}
\newcommand{\erg}{{~\rm erg}}
\newcommand{\yr}{{~\rm yr}}

\newcommand{\kpc}{{~\rm kpc}}

\begin{document}

\title{Natal kick by early-asymmetrical pairs of jets to the neutron star of supernova remnant S147}
\date{June 2025}

\author[0000-0002-9444-9460]{Dmitry Shishkin}
\affiliation{Department of Physics, Technion - Israel Institute of Technology, Haifa, 3200003, Israel}
\email[show]{s.dmitry@campus.technion.ac.il}  
\correspondingauthor{Dmitry Shishkin}

\author{Ealeal Bear}
\affiliation{Department of Physics, Technion - Israel Institute of Technology, Haifa, 3200003, Israel}
\email[]{ealeal44@technion.ac.il}  

\author[0000-0003-0375-8987]{Noam Soker}
\affiliation{Department of Physics, Technion - Israel Institute of Technology, Haifa, 3200003, Israel}
\email[]{soker@physics.technion.ac.il}  

\begin{abstract}
We analyze the bipolar morphology of the jet-shaped core-collapse supernova (CCSN) remnant (CCSNR) S147 and its neutron star (NS) kick velocity, and suggest that two pairs of unequal, opposite jets contributed to the NS kick velocity. This kick by early asymmetrical pairs (kick-BEAP) of jets mechanism operates within the framework of the jittering jets explosion mechanism (JJEM). We examine the prominent pair of large ears and, based on their flat structure rather than the more common conical structure of ears, conclude that two pairs of jets close in angle inflated the two opposite ears. We connect two opposite X-ray bright zones by an additional axis to create the full point-symmetric morphology of CCSNR S147. We propose that the two unequal jets that formed the X-ray bright zones imparted the first kick-BEAP, while the two pairs of jets that formed the ears imparted the second kick-BEAP. The two kick velocities are of about equal magnitude of $\simeq 450 \km \s^{-1}$, which implies very energetic jets. Such jets can excite gravitational waves that present detectors can detect from the Galaxy and the Magellanic Clouds. We use the morphology we identify to estimate the CCSNR age at $23,000 \yr$.  Our results strengthen the JJEM.
\end{abstract}



\section{Introduction}
\label{sec:Introduction}

\subsection{Two alternative explosion mechanisms}
\label{subsec:TwoMechanisms}

In recent years, studies have heavily referred to two theoretical alternative core-collapse supernova (CCSN) explosion mechanisms: the delayed neutrino explosion mechanism and the jittering-jets explosion mechanism (JJEM). 
In the delayed neutrino explosion mechanism of CCSNe, neutrinos that the cooling newly born neutron star (NS) emit, heat the gas in the `gain region', leading to the explosion (e.g., \citealt{Andresenetal2024, BoccioliFragione2024, Burrowsetal2024kick, JankaKresse2024, vanBaaletal2024, WangBurrows2024, Bambaetal2025CasA, Bocciolietal2025, EggenbergerAndersenetal2025, Huangetal2025, Imashevaetal2025, Janka2025, Laplaceetal2025, Maltsevetal2025, Maunderetal2025, Morietal2025, Mulleretal2025, Nakamuraetal2025, SykesMuller2025,  ParadisoCoughlin2025, Vinketal2025, WangBurrows2025}, for some very recent studies supporting the neutrino-driven mechanism). 
The gain region is immediately inner to the stalled shock at $r \simeq 100-150 \km$. Strong turbulence and instabilities in the gain region are crucial to a successful explosion in the neutrino-driven mechanism (e.g., \citealt{Buelletetal2023, Bocciolietal2025}). The magnetorotational explosion mechanism involves the launching of jets along a fixed axis. Therefore, it requires a rapidly rotating pre-collapse core to launch a pair of jets along a fixed axis (e.g., \citealt{Shibagakietal2024, ZhaMullerPowell2024, Shibataetal2025}, for recent studies of this mechanism). The demand for rapid pre-collapse rotation implies that this mechanism is rare; therefore, this mechanism attributes most CCSNe to the neutrino-driven mechanism. We, therefore, consider the magnetorotational explosion mechanism part of the neutrino-driven mechanism. 

According to the JJEM, jets explode all CCSNe. In most cases, these are jittering jets. In rare cases, where the core rapidly rotates, the jets are along a fixed axis; some fixed-axis explosions end with a black hole remnant.
The exploding, jittering jets come in pairs of opposite jets with fully or partially stochastic direction variations. Intermittent accretion disks or belts around the newly born NS, or a black hole if the NS collapses into a black hole, launch $N_{\rm 2j} \simeq 5-30$ pairs of jets on a time scale of $\tau_{\rm ex} \approx 0.5-10 \s$ (e.g., \citealt{Soker2025Learning}); later jets are possible. The jet-launching process might start as early as $\lesssim 0.1 \s$ after shock bounce, and the two jets might substantially differ in their power and opening angle \citep{Bearetal2025Puppis}; shock bounce is when the shock formation at the center occurs. 
The source of the stochastic angular momentum of accreted material is instabilities above the NS, seeded by angular momentum fluctuations due to the convection in collapsing core material (e.g., \citealt{GilkisSoker2014, GilkisSoker2016, ShishkinSoker2021, ShishkinSoker2023, WangShishkinSoker2024}). These instabilities below the stalled shock at $\simeq 150 \km$ include strong convection and turbulence (e.g., \citealt{Abdikamalovetal2016, KazeroniAbdikamalov2020}) and the spiral standing accretion shock instability (e.g., \citealt{Buelletetal2023} on the spiral mode of this instability). Although neutrino heating plays a role in adding energy to the jets \citep{Soker2022nu}, jittering jets are the primary driver of the explosion process and supply most of the explosion energy, even when an energetic magnetar supplies extra post-explosion energy (e.g., \citealt{SokerGilkis2017, Kumar2025}).

\citet{Soker2024UnivReview} summarized the status of the debate between the two explosion mechanisms at the end of 2024, \added{with a more recent update in \cite{Soker2025Learning}.} The neutrino-driven mechanism encounters some small to moderate problems and one major difficulty: (1) Not supplying the explosion energy of energetic CCSNe with explosion energies of $\gtrsim 2 \times 10^{51} \erg$, as some CCSNE are observed to have (e.g., \citealt{Moriyaetal2025}); (2) Predicting NS mass distribution that is displaced to somewhat higher NS masses than observed (\citealt{Soker2024UnivReview}). \added{Although the neutrino-driven simulations do produce NSs of low masses from stellar models with the initial minimum possible mass for CCSNe (e.g., \citealt{Muller_etal_2025_lowMassNS}), these models yield low explosion energies, i.e., $\ll 10^{51} \erg$, and account for a small fraction of all NS remnants. The distribution of NS masses that result from typical explosion energies yields NS mass distribution more massive than observed (\citealt{Soker2024UnivReview});} (3) Predicting failed CCSNe. Recent studies suggest that there is only a small population of failed CCSNe, or not at all (e.g., \citealt{ByrneFraser2022, StrotjohannOfekGalYam2024, BeasoretalLuminosity2025, Healyetal2025}). It might be that the neutrino-driven mechanism can overcome this difficulty, as \citet{Bocciolietal2025} suggested recently. 
(4) A major problem with the neutrino-driven mechanism is explaining the point-symmetric morphologies of CCSN remnants (CCSNRs); this problem likely rules out the neutrino-driven mechanism as the primary explosion mechanism.\footnote{Another noticeable problem of the neutrino-driven mechanism, though, is the qualitative and quantitative disagreements between simulations of different research groups (e.g., \cite{Janka2025}).}  On the other hand, the JJEM predicts that many CCSNRs exhibit point-symmetric morphologies, and new hydrodynamical simulations demonstrate this \citep{Braudoetal2025}.
The JJEM does not encounter these problems.
\added{ Point symmetric morphologies are defined to have two or more pairs of opposite structural features that do not share the same axis through the center. The opposite structural features include clumps, filaments, bubbles (faint structures closed and encircled by a brighter rim), lobes (bubbles with partial rims), nozzles (narrow openings in a rim), and ears (protrusions from the main CCSNR shell with decreasing cross sections away from the center).  }

\added{ On the other hand, the JJEM does have some challenges, and simulations of the neutrino-driven mechanism have some successes. Neutrino-driven simulations in three dimensions have demonstrated the explosion of some stellar models, but not all of them, and reproduced some observed properties. We note that some observed properties and outcomes are similar in the two explosion mechanisms \citep{Soker2025Learning}. The main achievement of simulations of the neutrino-driven mechanism is the robust demonstration that neutrino heating cannot be ignored in the explosion process. In the JJEM, neutrino heating boosts the jet-driven explosion \citep{Soker2022nu}. The challenges of the JJEM are that simulations do not reproduce stochastic accretion disks or jittering jets. One possible explanation \citep{Soker2025Learning} is that existing simulations and numerical codes lack the required resolution to accurately follow magnetic field processes, particularly reconnection, which are crucial for launching jets. Overall, the community should study the two alternative theoretical explosion mechanisms on an equal level and consider the advantages and challenges of both mechanisms. 
 }

\subsection{Bipolar and point-symmetry in CCSNRs}
\label{subsec:PointSymmetry}

According to the JJEM, launching pairs of opposite jets might shape pairs of structural features in the CCSNR. In general, the jets are not relativistic, as most CCSNe do not have relativistic jets, e.g., \citet{Guettaetal2020}; in the JJEM, relativistic jets, as in gamma ray bursts (e.g., \citealt{Izzoetal2019, AbdikamalovBeniamini2025}), might be a part of more pairs of jets that explode the star and power the supernova. Since the symmetry axis of the jets in the JJEM changes from one pair to the next, the pairs of structural features in the CCSNR will not share the same axis. This is a point symmetric morphology. Not all pairs of jets leave morphological imprints, as some are choked deep in the core and are too weak to leave observable marks. Only a fraction of the $N_{\rm 2j} \simeq 5-30$ pairs of exploding jets will leave observable marks. Instabilities in the explosion process, interaction with a circumstellar material (CSM) that the progenitor of the CCSN has lost before explosion (e.g., \citealt{Chiotellisetal2021, ChiotellisZapartasMeyer2024, Velazquezetal2023, Meyeretal2022, MeyerDetal2024}), and interaction with the interstellar medium (e.g.,  \citealt{Wuetal2019, YanLuetal2020, LuYanetal2021, MeyerMelianietal2024}) cannot form point-symmetric morphologies as \citet{SokerShishkin2025Vela} argued.
On the contrary, these processes, as well as an NS natal kick, a pulsar wind nebula (PWN), and heating processes such as reverse shock and radioactive decay, are likely to smear the point-symmetrical morphology. 

The following is a list of CCSNRs and the studies that attributed their point-symmetric morphologies to the JJEM: 
SNR 0540-69.3 \citep{Soker2022SNR0540},
CTB~1 \citep{BearSoker2023RNAAS}, 
the Vela CCSNR (\citealt{Soker2023SNRclass, SokerShishkin2025Vela}), 
the Cygnus Loop \citep{ShishkinKayeSoker2024},
N63A \citep{Soker2024CounterJet},
SN 1987A \citep{Soker2024NA1987A, Soker2024Keyhole}, 
G321.3–3.9 \citep{Soker2024CF, ShishkinSoker2025G321},
G107.7-5.1 \citep{Soker2024CF},
W44 \citep{Soker2024W44}, 
Cassiopeia A \citep{BearSoker2025}, 
Puppis A \citep{Bearetal2025Puppis},  
SNR G0.9+0.1 \citep{Soker2025G0901}, 
N132D \citep{Soker2025N132D}, 
Circinus X-1 \citep{SokerAkashi2025CirX1}, 
and the Crab Nebula \citep{ShishkinSoker2025Crab, ShishkinSoker2025Crab2}.

In cases where one pair of jets is significantly more energetic than the others and plays a major role in shaping the ejecta, the CCSNR exhibits a bipolar large-scale structure. Namely, there is only one pair of two opposite large lobes, bubbles, or ears. In this study, we examine images of SNR S147, which has one prominent pair of ears. We relate the asymmetrical ears to unequal jets in the pair of jets that shaped the two ears. We suggest that pairs of unequal jets impart a natal kick velocity to the NS of SNR S147. This is the kick by the early asymmetrical pair (kick-BEAP) mechanism. 

A comment is in place here. We apply visual inspection to classify point-symmetric and bipolar CCSNRs. This method of visual inspection classification of morphologies is very common in classifying planetary nebulae (e.g., \citealt{Balick1987, Chuetal1987, Sahaietal2007, Sahaietal2011}), and AGN jets (e.g., \citealt{Hortonetal2025}); it is a powerful tool that has led to significant breakthroughs, e.g., establishing shaping of planetary nebulae by binary interaction and the major role of jets that the companion launches. For example, visual inspection can differentiate between CCSNe and SNe Ia in most cases, a capability that some other mathematical tools, such as the new multipole moment analysis method for SNR X-ray images with a circular boundary \citep{Leahyetal2025}, lack. 

\subsection{The kick-BEAP mechanism for NS kick velocity}
\label{subsec:kickBeap}

There are several theoretical mechanisms to impart kick velocity to the NS at CCSN explosion (e.g., \citealt{YamasakiFoglizzo2008, Igoshev2020, Yao_etal_2021_3Dspinkick, Xuetal2022, LambiasePoddar2025}). Early JJEM studies assumed that the main mechanism to impart a kick velocity is the tug-boat mechanism, which was developed in the frame of the neutrino-driven explosion mechanism (e.g., \citealt{Schecketal2004, Schecketal2006, Nordhausetal2010, Nordhausetal2012, Wongwathanaratetal2010, Wongwathanaratetal2013kick, Janka2017}). In the tug-boat mechanism, a massive clump that is ejected from near the center, gravitationally pulls the NS along its direction. More recently, we proposed the kick by early asymmetrical pairs (kick-BEAP) mechanism \citep{Bearetal2025Puppis}. In the kick-BEAP mechanism, which operates in the JJEM, one to a few pairs of energetic jets are produced with unequal jets. In each pair, one jet is significantly more powerful than the other, carrying more momentum. From momentum conservation, the NS acquires velocity in the opposite direction to the more powerful jet. Additionally, from momentum conservation, a massive ejecta and/or faster ejecta expand in the opposite direction of the NS in both mechanisms.

\added{What makes the kick-BEAP possible is the stochastic nature of angular momentum and the turbulence above the NS, which results in short accretion episodes. The typical accretion period of $\simeq 0.01-0.1 \s$ is not much longer, or even shorter, than the relaxation time of the accretion disk, such that the disk has no time to fully relax \citep{Bearetal2025Puppis}. Since accretion does not start symmetrically due to turbulence and angular momentum fluctuations, the two sides of the disk do not begin equally and have no time to relax. The unequal sides of the accretion disk are likely to launch unequal jets, in terms of their power and opening angle. One jet might be weaker than the other, and momentum conservation implies an NS kick towards the weak jet. The magnitude of the effect can explain observed NS kicks (\citealt{Bearetal2025Puppis} and section \ref{subsec:KickBeapS147}). Although it can work, the kick-BEAP is less efficient in the magnetorotational mechanism because, in this mechanism, rapid pre-collapse core rotation leads to a long-lived accretion disk, $\gg 0.1 \s$. This implies that the accretion disk launching the fixed-axis jets lives much longer than the relaxation time and has time to relax. The two sides of the disk are equal and likely to launch equal jets. Only at the very early phase of the accretion disk, before it relaxes, the kick-BEAP might operate efficiently.} 

Observations show anti-correlation between the primary X-ray emitting ejecta and the neutron star velocity in Cassiopeia A (e.g., \citealt{HwangLaming2012}). 
This is compatible with both the tug-boat mechanism and the kick-BEAP mechanism. The asymmetrical ejecta of Puppis A was our motivation in \citet{Bearetal2025Puppis} to propose the kick-BEAP mechanism. In the JJEM, both mechanisms can operate: the kick-BEAP earlier in the explosion, while the tug-boat mechanism operates at relatively later times, $\simeq 0.5-5 \s$ (e.g., \citealt{Wongwathanaratetal2013kick}). 

Spin-kick alignment is observed in many pulsars (e.g., \citealt{Johnstonetal2005, Noutsosetal2012, BiryukovBeskin2025}), but not all. The best example is the NS in the CCSNR S147 ($\rm J0538+2817$, e.g., \citealt{BiryukovBeskin2025}). The kick-BEAP explains this alignment as follows: the kick velocity is along the angular momentum axis of the material that fed the disk, which launched the two unequal jets; the same material spun up the NS with a spin axis aligned along the same angular momentum axis. It is unclear whether the neutrino-driven mechanism can explain this alignment.     

The kick-BEAP mechanism can account for the new findings of \citet{Vallietal2025} who thoroughly analyzed 23 Be X-ray binaries, i.e., with an NS accreting mass from a Be star. Fitting progenitors to the binary eccentricity and orbital period, they identified two populations of NS according to their natal kick velocity: one of NSs with natal kick velocities $< 10 \km \s^{-1}$ and one with natal kick velocities around $100 \km \s^{-1}$ and aligned within $5^\circ$ from the binary angular momentum. The explosions that formed these NSs were stripped-envelope CCSNe, implying that the progenitors lost large amounts of mass. The mass loss was due to the binary interaction that likely spun up the progenitor (e.g., \citealt{Gilkis_etal_2025_binaryCCSN_WR}). 
According to the JJEM, the explosion launched one or more powerful pairs of unequal jets along the angular momentum axis. This unequal pair imparted the natal kick to the NS. We find the new findings of \citet{Vallietal2025} to support the kick-BEAP mechanism in the frame of the JJEM. 
 
The distribution of angles between the NS kick velocity and the jet-main axis (which is not always along the spin axis) tends to avoid small angles \citep{BearSoker2018kick, BearSoker2023RNAAS,  Soker2022SNR0540}. Within the framework of the tug-boat mechanism, this implies that the main-jet axis and the dense clump that accelerated the NS do not align. There are two possible explanations. (1) The jets prevent the formation of dense clumps along their propagation directions. Therefore, no NS acceleration takes place in those directions. (2) The dense clump that pulled the NS with it and imparted the NS the kick velocity is part of a group of clumps. Some clumps feed the NS to launch jets. The angular momentum is perpendicular to this direction of motion. (3) The third explanation holds for the tug-boat and kick-BEAP mechanisms. After the NS acquired its kick velocity, it accretes mass with angular momentum at a large angle to its motion, as part of the relative angular momentum is due to the NS velocity. The accreted mass launches jets along its angular momentum axis, which is at a large angle to the kick velocity. This last possibility seems more general and might account for most cases of small-angle avoidance.  

In this study, we apply the JJEM to SNR S147 (Semeis 147) and discuss the relation between its bipolar morphology and its kick velocity (Section \ref{sec:S147}). In Section \ref{sec:GW}, we discuss possible implications to the detection of gravitational waves from CCSNe. 
We summarize in Section \ref{sec:Summary}. 

\section{NS kick in two episodes in SNR S147}
\label{sec:S147}

\subsection{The general morphology of SNR S147}
\label{subsec:MorphologyS147}

The CCSNR S147 \citep{GazeShajn1952} presents an inhomogeneous structure in the X-ray (e.g., \citealt{Michailidis_etal_2024_S147Xray, Khabibullinetal2024_Xray}), radio (e.g., \citealt{Xiaoetal2008}, \citealt{Khabibullinetal2024_Radio}), and visible (e.g., \citealt{Drewetal2005, Renetal2018RAA}).  
In H$\alpha$ emission, e.g., \citet{Lozinskaia1976}, it presents many filaments (hence termed the “Spaghetti Nebula”). 
\citet{Dinceletal2015} discovered an OB runaway star inside S147, and suggested the progenitor was in a binary system.
CCSNR S147 poses some puzzles, like the kinematics of the NS (PSR J0538+2817) that suggests a younger system age than the estimated age of the SNR, e.g., \citet{Khabibullinetal2024_Xray}, who also noted that an expansion of the SNR inside a low-density wind-blown bubble (e.g., \citealt{Hughes1987}) might solve the age discrepancy.
We focus on the relationship between several directions defined by SNR S147: the main jet axis, which we find to likely be two pairs of jets, the line connecting the two X-ray bright regions, and the NS's proper motion direction. 
In the Appendix, we use the morphology we identify to estimate the age of S147. 

\begin{figure}[th!]
\begin{center}
\includegraphics[trim=0.14cm 0.1cm 0cm 0.3cm, clip, width=0.51\textwidth]{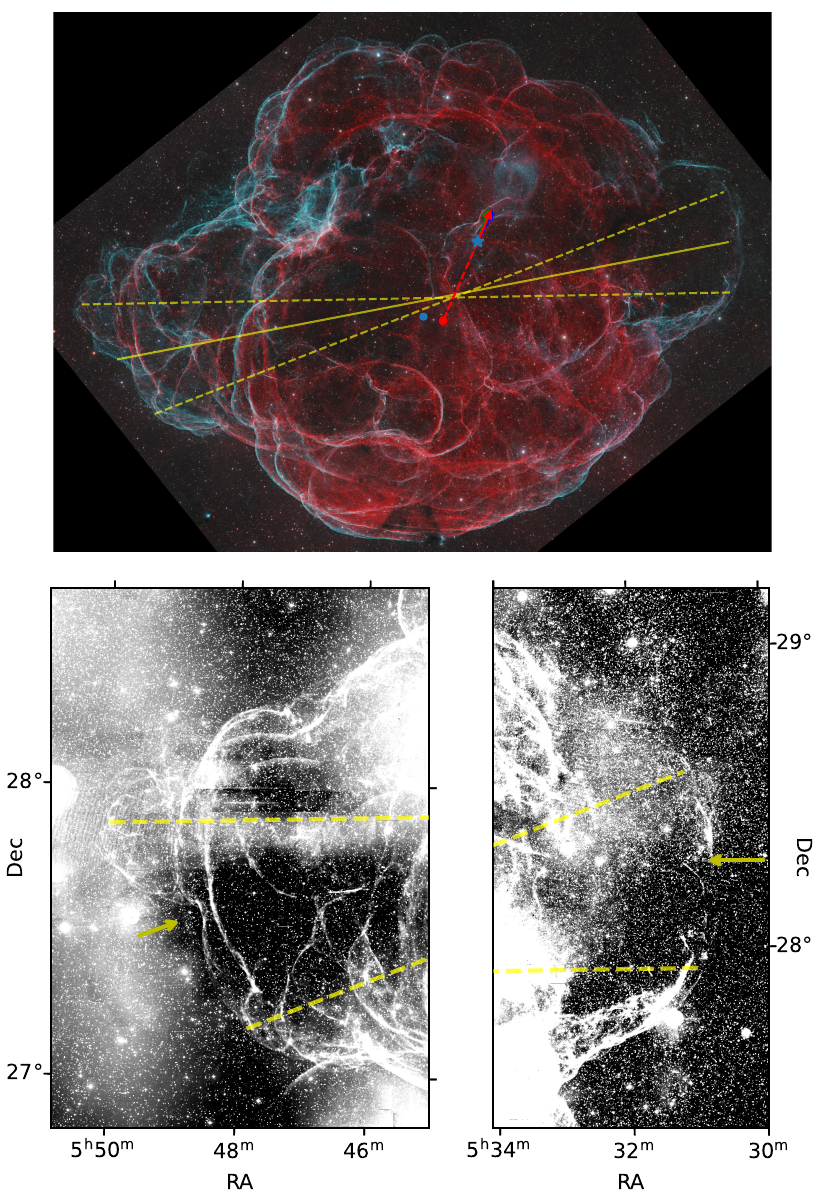} 
\caption{\textbf{Upper panel:} An Image of SNR S147 taken and processed by Mr. Christian Koll (\url{https://app.astrobin.com/i/vh6kx6}). The blue and green colors represent [O \textsc{iii}] and the red represents H$\alpha$. The two dashed yellow lines represent our proposed symmetry axes for two possible pairs of jets that shaped the two wide ears (see text). We draw these two symmetry axes by the sub-structures in each wide ear, so that they cross at the path continuation of the proper motion of the NS (dashed-red line and red arrow; See Figure~\ref{fig:S147} for details). We mark the location of the NS with a blue star. The blue dot is the geometric center of S147 as suggested by \citet{Kramer_etal_2003_center} based on radio observations. We draw a yellow solid line between the two axes, which represents the main jet axis of S147. \textbf{Lower panels:} Over-exposed H$\alpha$ close-ups on the ear structures, taken from a mosaic image composed of IPHAS DR2 survey data (see text for details). White smeared regions are artifacts of the over-exposure. The filamentary structure is nevertheless clear.} 
\label{fig:S147morphology}
\end{center}
\end{figure}

The [O \textsc{iii}] and H$\alpha$ images of S147 reveal two opposite wide ears, with a filamentary structure that extends into the ears. We present a composite image of these two emission lines in the upper panel of Figure \ref{fig:S147morphology}, adapted from Mr. Christian Koll. Some earlier studies have connected these two ears with a single line, roughly through the center of the SNR (e.g., \citealt{Gvaramadze2006, GrichenerSoker2017}).
In the lower panels of Fig~\ref{fig:S147morphology} we present a close-up of the H$\alpha$ filamentary structure of the two ears at the edges of S147, from a mosaic within the combined IGAPS survey \citep{Greimel_etal_2021_IGAPS}, available on the data website\footnote{http://www.star.ucl.ac.uk/IGAPS/data.shtml} as imaged in the IPHAS DR2 \citep{Barentsen_etal_2014_IPHASDR2}. We overexpose the images to highlight the sub-structures that define the protrusions and their separation into two pairs. We mark the same dashed lines as in the upper panel for reference.

Motivated by identifying point-symmetric morphologies in over ten CCSNRs (Section \ref{subsec:PointSymmetry}), we carefully examine for a possible point-symmetric morphology in CCSNR S147.

We note the following prominent morphological features of CCSNR S147. 
\begin{enumerate}
    \item The east (left in the figure) ear is larger and brighter than the west ear. 
    \item The ears are wide and do not have the conical structure typical of many ears. 
    \item The eastern ear protrudes further to the east from its northern region.  
    \item The west ear has two bright arcs on its boundary with an almost dark zone in between.
    \item A slight indentation can be identified at the approximate middle point of each ear (yellow arrows in the lower panels of Fig~\ref{fig:S147morphology}).
\end{enumerate}
Based on properties 2-5, we propose that two pairs of jets shaped the two ears. We take the axis of one pair of jets from the protrusion of the east ear to the bright rim on the south of the west ear, and the other axis from the center of the south of the east ear that has no protrusion to the north rim of the west ear. We mark these two axes by yellow-dashed lines on Figure \ref{fig:S147morphology}. The locations of the ends of the axes are not well defined and serve to illustrate the different axes of symmetry. We ensure that the two axes cross along the trajectory of the NS (dashed-red line) according to its' proper motion direction, the red arrow in Figure \ref{fig:S147morphology}.

By examining the possible intersection points between the lines connecting opposite ears from each pair, we can estimate the distribution of intersection points along the NS trajectory. 
We derive the trajectory using \added{the \texttt{Astropy} python package based on }the proper motion estimates of \citet{Chatterjee_etal_2009_NSpm}: $\mu_\alpha^*=-23.57^{+0.10}_{-0.10}\ \frac{\rm{mas}}{\rm{yr}}, \mu_\delta=52.87^{+0.09}_{-0.10}\ \frac{\rm{mas}}{\rm{yr}}$ \added{who fit parallax and proper motion to multi-epoch VLBA observations of NS} 
(see also \citealt{Ng_etal_2007_NSpm}). The intersection point is the origin of the NS and jets (and hence, explosion), and corresponds to an age of $\tau_{S147}=23.2^{+2.2}_{-2.5}\ \rm kyr$, where the bounds originate from the distribution of points around the mean intersection point. We elaborate on this analysis, \added{uncertainties}, our age estimate and previous age estimates in Appendix~\ref{appendix:Appendix_Age}.

These two pairs of jets, which we claim shaped the ears, had similar energy and were at a small angle to each other. 
We assume that the newly born NS launched these two pairs one after the other, or that there was one precessing pair instead of two pairs of jets. In any case, we consider the two pairs of jets as a single NS kick velocity episode for our study of the kick velocity episodes to follow. During this kick episode, we take the net kick velocity direction along the solid-yellow line (defined by connecting the midpoints of each pair) that we draw on the upper panel of Figure \ref{fig:S147morphology}. 

Our main finding in this section is that the ears of CCSNR S147 reveal a point-symmetric morphology, i.e., two pairs of jets shape the ears. By examining where this symmetry axis falls along the NS trajectory, we estimate the age of S147 to be $\tau_{S147}=23.2^{+2.2}_{-2.5}\ \rm kyr$

\subsection{The NS kick episodes}
\label{subsec:KickEpisodes}

In Section \ref{subsec:MorphologyS147} we identify two pairs of jets that shaped the ears (yellow-dashed lines on Figure \ref{fig:S147morphology}). To study the kick velocity, we take the main jet axis of S147 to be between these two axes and more or less through the center of the two ears (the yellow-solid line), and, in the frame of the kick-BEAP mechanism, consider the two pairs of jets as one kick episode along the main jet axis. We turn to analyze X-ray observations of S147 for an additional symmetry axis.

\citet{Michailidis_etal_2024_S147Xray} used the SRG eROSITA X-ray observations \citep{Predehl_eROSITA_2021, Merloni_eROSITA_2024} to image CCSNR S147 and analyzed individual sub-regions to extract spectral parameters, including metal composition. We use the extracted images and spectral analysis of \citet{Michailidis_etal_2024_S147Xray} and the same observations (see observations details within) to examine the diffuse X-ray emission of S147. 

\begin{figure*}
\begin{center}
\includegraphics[trim=0.5cm 2.8cm 0.5cm 2.8cm, clip, width=\textwidth]{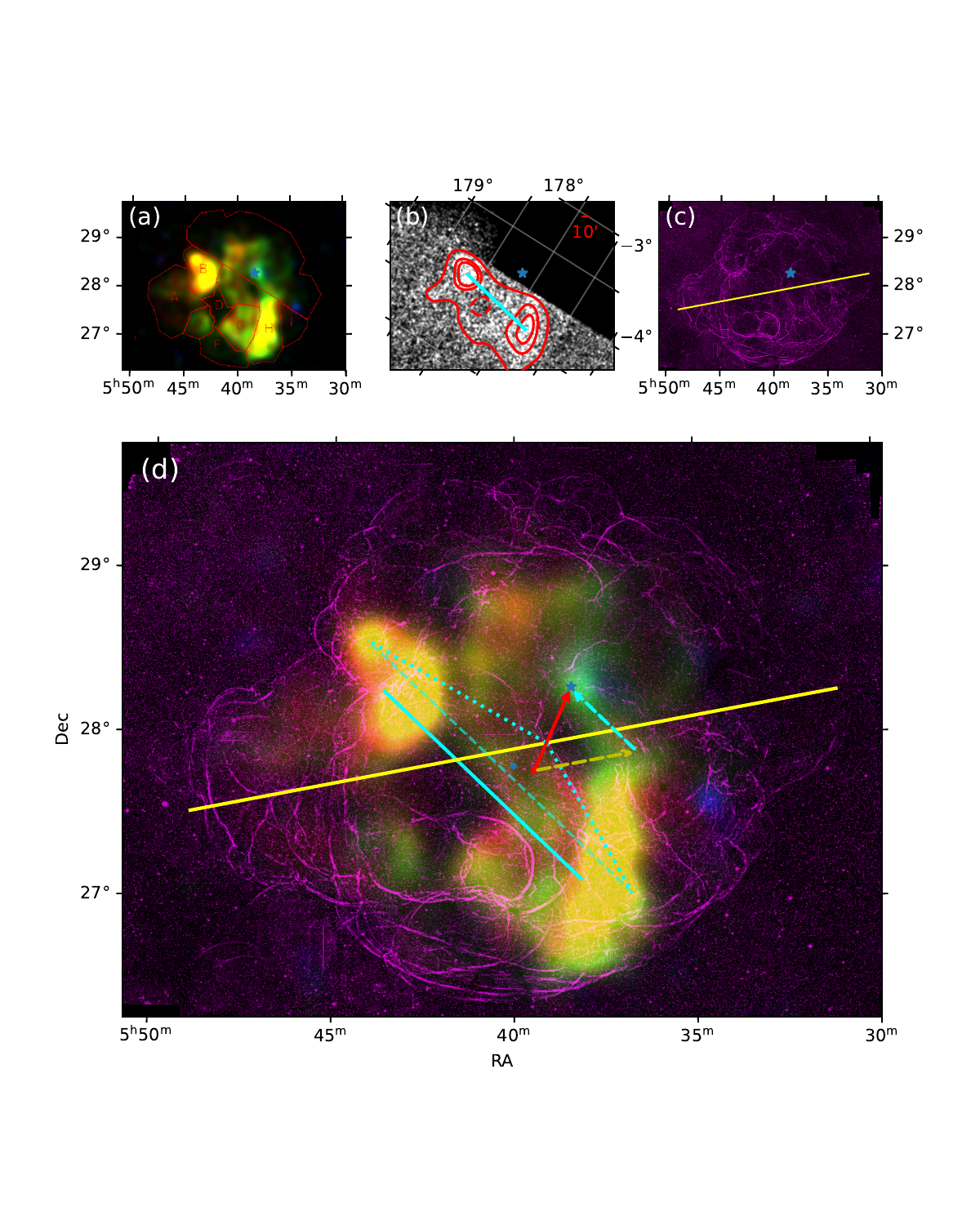} 
\caption{An annotated view of S147, combining eROSITA X-ray and IGAPS H$\alpha$. \textbf{(a)} RGB X-ray image, spanning $0.3-0.6\rm\ keV$ (red), $0.6-1.0\rm\ keV$ (green), $1.0-1.5\rm\ keV$ (blue) from \citet{Michailidis_etal_2024_S147Xray} with denoted sub-regions they used to perform spectral analysis. Note the surface-bright regions B and H. We denote the current NS location with a blue asterisk in this panel and all subsequent panels. \textbf{(b)} X-ray counts image of S147 from eROSITA DR1 data in the $0.2-2.3\rm\ keV$ range, point-source removed \citep{ShishkinSoker2025G321} and gaussian smoothed to $1'$ scale. Red contours (three arbitrary levels) are of a $10'$ smoothed image (see scale bar in the upper right), revealing two main peaks in the diffuse X-ray emission map. A Cyan line connects the centers of the two top contours. Note the negative-area contour (dashed contour) coinciding with region D (panel (a)). \textbf{(c)} IGAPS H$\alpha$ mosaic image \citep{Greimel_etal_2021_IGAPS}, processed (morphologically filtered using $\rm OpenCV$; \citealt{opencv_library}) to enhance the filamentary structure of S147. A yellow line denotes the main axis of S147, connecting two prominent ear structures on opposite ends. \textbf{(d)} A combined X-ray and H$\alpha$ image of S147, with the focal-connecting cyan line from panel (b) and the ear-axis from panel (c). A red arrow denotes the NS trajectory according to the proper-motion direction of the NS \citep{Chatterjee_etal_2009_NSpm}, starting from an assumed explosion age of $35\rm\ kyr$ \citep{Yao_etal_2021_3Dspinkick}. A red dot denotes the proposed origin, and a blue dot marks the radio geometric center of S147 as denoted in \citet{Kramer_etal_2003_center}. From the proposed origin, we project the contribution of each kick direction with dashed arrows (cyan for the X-ray, yellow for the asymmetrical ears). The ratio between the lengths of the two dashed arrows is $D_{\rm X-ray}=0.88\ D_{\rm ears}$, corresponding to a slightly different kick magnitude in the two kick episodes  (Sec~\ref{subsec:KickBeapS147}).}
\label{fig:S147}
\end{center}
\end{figure*}

In Fig~\ref{fig:S147}, we present annotated images of S147 in X-ray and H$\alpha$ (See Fig~\ref{fig:S147morphology} and Sec~\ref{subsec:MorphologyS147} for details). All individual panels display the same patch of sky, with a blue star denoting the NS location.
In panel (a) of Figure~\ref{fig:S147} (Fig~\ref{fig:S147}a) we present an adapted version of the \citet{Michailidis_etal_2024_S147Xray} X-ray RGB image (R: $0.3-0.6\rm\ keV$, G: $0.6-1.0\rm\ keV$, B: $1.0-1.5\rm\ keV$) and their division into sub-regions. Two surface-bright regions B and H, and the metal (O, Ne, Mg) rich regions B, E, G, I (See \citealt{Michailidis_etal_2024_S147Xray} for spectral parameters) form a symmetry axis, which we draw in Figure \ref{fig:S147}b.

In Figure~\ref{fig:S147}b we present an X-ray count image in the $0.2 - 2.3\rm\ keV$ range of the publicly available eRASS1 DR1 data, around the S147 remnant. Note the upper region (corresponding to region J in Fig~\ref{fig:S147}a) is missing, as S147 is at the edge of the eRASS1 DR1 release domain ($l>180$). The NS resides in the missing region. We use eSASS4DR1 \citep{Brunner_eSASS_2022} to combine, reproject, and merge the relevant event maps. Point sources (eRASS1 main catalog, \citealt{Merloni_eROSITA_2024}) were removed with dedicated software designed to retain diffuse emission \citep{ShishkinSoker2025G321}. The image was smoothed on two scales: $1'$ for display purposes and $10'$ for the contour lines, which are depicted in red lines in the panel (solid for hill contours and dashed for depression contours). The contour lines for the smoothed image reveal two main centers (or peaks) for the X-ray emission, which we connect with a solid cyan line to denote another symmetry axis; we argue for a second NS kick episode along this axis.

In Figure~\ref{fig:S147}c we show a processed version of the H$\alpha$ mosaic from the IGAPS combined survey \citep{Greimel_etal_2021_IGAPS}. We pass the original mosaic through a filter (`Top-hat' morphological transformation with a 20x20 pixel kernel, as implemented in $\rm OpenCV$; \citealt{opencv_library}) to strengthen sharp edges, as to enhance the filamentary structure defining the general shape and edges of S147. We mark with a yellow-solid line the approximate symmetry axis of the two ears, as we did in Figure~\ref{fig:S147morphology}.

In Figure~\ref{fig:S147}d we combine the X-ray (RGB, \citealt{Michailidis_etal_2024_S147Xray}) and H$\alpha$ \citep{Greimel_etal_2021_IGAPS} images. We mark the axes from panels (b) and (c).
With a dashed cyan line we draw a shifted X-ray axis in the direction of the NS to account for the missing counts from region J (panels (a), (b)), so it starts at the far end of the bright X-ray emission at the north-east (panels (a), (d)). 
We extend the proper motion \citep{Chatterjee_etal_2009_NSpm} direction to form a trajectory and an origin, assuming an age of $35\rm\ kyr$ \citep{Yao_etal_2021_3Dspinkick}, where we mark the trajectory with a solid red line and the origin with a red dot. For reference, a blue dot marks the geometric center as derived in \citet{Kramer_etal_2003_center}. With a broken dotted cyan line, we connect the edges of the shifted X-ray axis to the intersection point of the ear-axis and the NS trajectory, and produce a ``bent axis'' that signifies an asymmetric pair of jets. Namely, the two opposite jets were at an angle of $150^\circ$ rather than $180^\circ$. 
We take the directions defined by the two axes (X-ray - cyan, ear-pair - yellow) to be the direction of two NS kick episodes, as marked by the corresponding dashed arrows, and build a vector sum to give the final NS kick velocity (the red-solid arrow). The ratio of the lengths of the two arrows is $D_{\rm X-ray}/ D_{\rm ears}=0.88$, and the angle between them is $\alpha_{\rm  X-ray}- \alpha_{\rm ears} = 136^\circ - 10^\circ = 126^\circ$. 

For a NS velocity of $407^{+79}_{-57}\ \rm{km\ s^{-1}}$ \citep{Yao_etal_2021_3Dspinkick} in S147 , we find that, for the triangle we draw on Figure~\ref{fig:S147}d, the NS kick velocity along the axis through ears and through the X-ray bright zones should be 
\begin{equation}
\Delta v_{\rm NS,E} \simeq 473^{+92}_{-66} \km \s^{-1}, \qquad \Delta v_{\rm NS,X} \simeq 416^{+81}_{-58} \km \s^{-1},   
\label{eq:TwoKicks}
\end{equation}
where the errors stem from the uncertainty in the NS velocity.

Our main claim in this section is that two main NS-kick episodes occurred during the explosion of the S147 progenitor, contributing to the total kick velocity and determining the kick direction.

\subsection{Applying the kick-BEAP mechanism to SNR S147}
\label{subsec:KickBeapS147}

We apply the kick-BEAP mechanism (\citealt{Bearetal2025Puppis}; Section \ref{subsec:kickBeap}) to the NS of SNR S147. Each of the two kick episodes we proposed in \ref{subsec:MorphologyS147} imparted a velocity of magnitude (equation \ref{eq:TwoKicks}) of $\Delta v_{{\rm NS}} \simeq 450 \km \s^{-1}$.  We consider episodes with high mass-accretion rates and accretion disks that launch massive, opposite, unequal jets at relatively low velocities, as in \citet{Bearetal2025Puppis}. The mass the NS accretes in a kick episode is $M_{\rm acc}$, the two opposite jets carry a fraction $f_{\rm j1}$ and $f_{\rm j2}$ of this mass respectively; the inequality $f_{\rm j2} \ll f_{\rm j1}$ holds for a high kick velocity. 
The kick velocity the jets impart and the jets' combined energy are  
\begin{equation}
\begin{split}
\Delta v_{\rm NS} & =  430
\left( \frac{f_{\rm j1}-f_{\rm j2}}{0.12} \right)  
\left( \frac{M_{\rm acc}}{0.1 M_\odot} \right) \\ & \times 
\left( \frac{v_{\rm j}}{5 \times 10^4 \km \s^{-1}} \right) 
\left( \frac{M_{\rm NS}}{1.4 M_\odot} \right)^{-1}
\km \s^{-1}, 
\label{eq:V1jet}
\end{split}
\end{equation}
and 
\begin{equation}
\begin{split}
E_{\rm j12} & =  5 \times 10^{50}    
\left( \frac{f_{\rm j1}+f_{\rm j2}}{0.2} \right)  
\left( \frac{M_{\rm acc}}{0.1 M_\odot} \right) \\ & \times 
\left( \frac{v_{\rm j}}{5 \times 10^4 \km \s^{-1}} \right)^2 
\erg, 
\label{eq:E1jet}
\end{split}
\end{equation}
respectively. In this scaling $f_{\rm j1}=0.16$ and $f_{\rm j2}=0.04$. i.e., one jet is four times as strong as the other.   

The conclusion is that two jet-launching episodes during the explosion of the progenitor of SNR S147 were very energetic and likely comprised most of the explosion energy.
\citet{GrichenerSoker2017} estimated the extra energy to inflate the two ears of S147 to be $\epsilon_{\rm east} = 0.072$ and $\epsilon_{\rm west} = 0.039$,   where $\epsilon$ is the fraction out of the total explosion energy. However, \citet{GrichenerSoker2017} referred only to the protrusions of the ears out from the main SNR shell. Our results suggest that the jets could have influenced zones within the SNR main shell. In equations (\ref{eq:V1jet})  and (\ref{eq:E1jet}) we scale with  $\epsilon_{\rm east} = 0.4/E_{\rm ex,51}$ and $\epsilon_{\rm west} = 0.1/E_{\rm ex,51}$, for an explosion energy of $E_{\rm ex} = 10^{51} E_{\rm ex,51} \erg$. For these parameters, an explosion energy of $E_{\rm ex,51} \simeq 1.5-2$ would better fit the large kick velocity. 

We showed in this section that the kick-BEAP mechanism can reasonably account for the kick velocity of the NS of S147. 

\section{Implications to gravitational waves}
\label{sec:GW}

\citet{Soker2023GW}  estimated the gravitational waves that the jet-inflated bubbles (cocoons) might radiate in the JJEM. The source of the gravitational wave is the turbulence in the jet-inflated bubbles; simulations show that the bubbles (cocoons) contain large vortices \citep{Braudoetal2025}. \citet{Gottliebetal2023} suggested and studied this gravitational-wave source for relativistic jets in black hole formation. \citet{Soker2023GW} estimated the expected properties of the gravitational waves from CCSNe according to the JJEM for pairs of jets that each pair carries an energy of $E_{\rm j12} = 10^{50} \erg$.  

\begin{figure} [t]
\centering
\includegraphics[trim=2.5cm 14.75cm 2.0cm 1.7cm ,clip, scale=0.52]{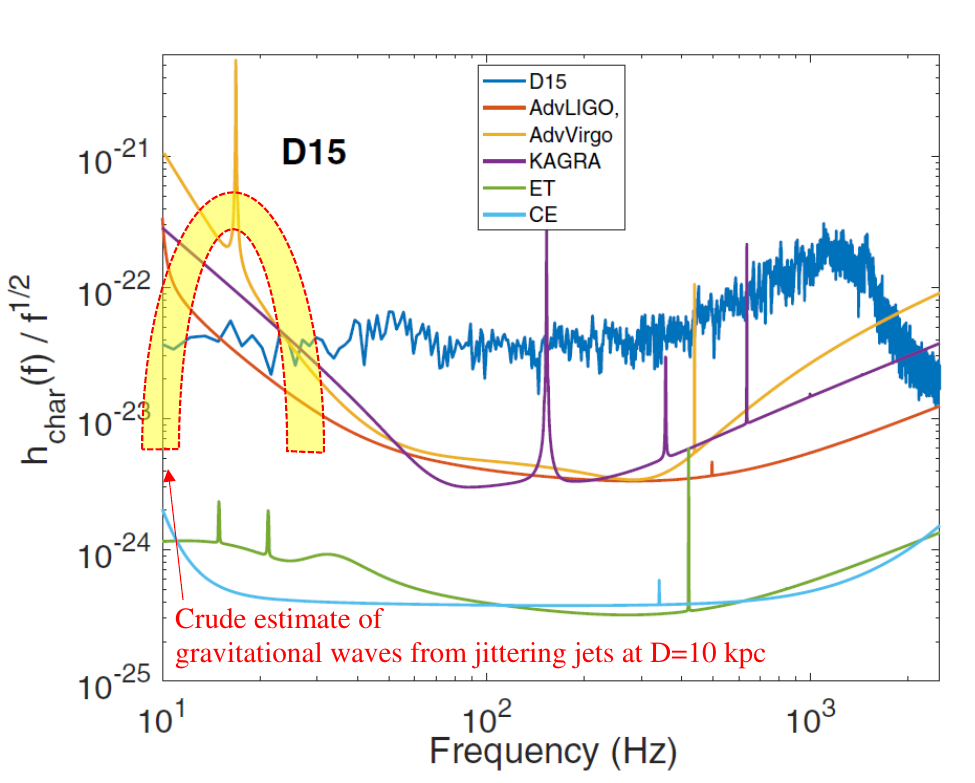}
\caption{A figure adapted from \citet{Soker2023GW}, who added a horseshoe-shaped yellow zone to a figure from \citet{Mezzacappaetal2023}. The horseshoe-shaped yellow zone here is higher than in \citet{Soker2023GW}. It is a crude estimate of the characteristic spectrum of $h f^{-1/2}$ by an energetic pair of jets that imparts a NS kick velocity of hundreds of $\kms$ in the frame of the JJEM of a CCSN at a distance of $D=10 \kpc$. The pair of jets carries an energy of $E_{\rm j12} \simeq 5 \times  10^{50} \erg$. 
We base the horseshoe-shaped yellow zone on \citet{Soker2023GW}.  
The blue `noisy' line, calculated by \citet{Mezzacappaetal2023}, is the characteristic gravitational wave strain from a CCSN of a $15 M_\odot$ stellar model in the frame of the neutrino-driven mechanism. The other colored lines represent the sensitivity of five gravitational wave detectors. The current-generation detectors are the Advanced Laser Interferometer Gravitational Observatory (AdvLIGO), the Advanced VIRGO, and the Kamioka Gravitational Wave Detector (KAGRA). The next-generation detectors, Cosmic Explorer and the Einstein Telescope, are more sensitive. The predicted observed full gravitational wave spectrum encompasses both the contributions from the regions near the NS, which exist in both the JJEM and the neutrino-driven explosion mechanism (blue line), and the contribution from the jittering jets (yellow horseshoe-shaped zone).   
} 
\label{Fig:StrainFrequency}
\end{figure}

The jet pairs that impart the NS kick velocity of SNR S147, and possibly some other CCSNe, are very energetic, and can contribute a fraction of $\simeq 0.25-0.5$ of the explosion energy. 
\added{  This fraction comes from our estimate of the energy that the two jets that imparted the NS kick velocity carry, equation (\ref{eq:E1jet}), and a typical CCSN explosion energy of $\simeq 10^{51} \erg - 2 \times 10^{51} \erg$. }
We, therefore, scale up by a factor of five the energy of the predicted gravitational wave signal from the most energetic jet-inflated bubbles in the JJEM. 
We present the results in Figure \ref{Fig:StrainFrequency}. We adapted a figure from \citet{Mezzacappaetal2023}, and added the predicted characteristic of the gravitational wave with a horseshoe-shaped yellow zone, scaled from \citet{Soker2023GW}.  The noisy blue line represents a gravitational wave prediction based on the neutrino-driven mechanism. As the turbulent motion near the newly born neutron star is similar in both explosion mechanisms, we predict the total gravitational wave signal to be the combination of the noisy blue line and the horseshoe-shaped yellow zone. 

Gravitational waves can teach us about some CCSN explosion properties (e.g., \citealt{Iessetal2023, Mitraetal2024}); hence, we should include all possible CCSN sources of gravitational waves.
\citet{Soker2023GW} concluded that detecting gravitational waves from jet-inflated bubbles is possible for a Galactic CCSN, but marginal. We conclude that a clear signal will be detected from a galactic CCSN, possibly even a CCSN in the Magellanic Clouds.

\section{Summary}
\label{sec:Summary}

We studied the CCSNR S147, which exhibits a bipolar morphology characterized by a prominent pair of large ears, and related its NS velocity to its morphology.
Based on the wide morphology of the ears rather than the more common conical structure and their internal structures, in Section \ref{subsec:MorphologyS147} we proposed that the ears were inflated by two pairs of jets (or a precessing pair of jets). We draw the two axes of the two pairs of jets on the upper panel of Figure \ref{fig:S147morphology} by dashed yellow lines. This point-symmetric morphology is completely compatible with the JJEM and challenges the neutrino-driven explosion mechanism. 

\added{Let us elaborate on the last important conclusion. According to the neutrino-driven simulations (see references in Section \ref{sec:Introduction}), pre-collapse fluctuations, like the convection in the collapsing core, and instabilities above the neutron star and behind the stalled shock, facilitate the revival of the stalled shock. Protruding `fingers' of energetic material from the gain region behind the stalled shock revive the stalled shock in several places and explode the core and the star. This process is prone to instabilities. As a result of the instabilities, the neutrino-driven explosion forms filaments and clumps in the CCSNR. However, in the neutrino-driven mechanism, \textit{these are distributed randomly rather than in pairs.} The distribution of clumps, filaments, bubbles, lobes, ears, and nuzzles in pairs, which observations reveal in 16 CCSNRs (see list of 15 CCSNRs in Section \ref{sec:Introduction}, to which we add S147), are the point-symmetric morphologies that the JJEM predicts. As far as we know, no study of the neutrino-driven mechanism suggested an explanation for point-symmetric morphologies. The magnetorotational explosion mechanism requires a rapid pre-collapse core rotation, implying a fixed-axis angular momentum. Consequently, the accretion disk does not jitter much, and the accretion disk launches the jets along a fixed axis. The magnetorotational explosion mechanism can form pairs of structural features, but only along the same axis; it cannot account for two or more axes in different directions. Namely, it cannot account for point-symmetric morphologies. Additionally, the magnetorotational mechanism is rare because it requires a rapidly rotating core before collapse.   }   

\added{ Overall, our results, together with other studies of point-symmetric CCSNRs in the last two years, put on solid ground our group's ten-year-old call (e.g., \citealt{Papishetal2015, Soker2017Magnetar}) for a paradigm shift from neutrino-driven to jet-driven CCSN explosion mechanism (\citealt{Piranetal2019} followed with a similar but weaker call). } 

To connect the NS kick velocity to the morphology of S147, we define one axis through the ears and between the two jet axes; this is the main jet axis of S147 (solid yellow line in Figures \ref{fig:S147morphology} and \ref{fig:S147}). 
The projected intersection between the ears symmetry axis and the NS trajectory allowed us to estimate the age of the NS-SNR system to be $\tau_{S147}=23.2^{+2.2}_{-2.5}\ \rm kyr$ (Section \ref{subsec:MorphologyS147} and Appendix~\ref{appendix:Appendix_Age}).

The X-ray image of S147 (Figure \ref{fig:S147}) reveals two X-ray bright zones. We connected them by an additional symmetry axis of S147 (cyan lines). We suggested that an early pair of unequal jets compressed the X-ray bright zones, and that two pairs of unequal jets, close in time and direction, shaped the two ears later in the explosion process. The unequal power of the opposite jets in these pairs of jets imparted a kick velocity to the NS in a direction opposite to that of the more powerful jet; this is the kick-BEAP mechanism. We draw the two kick directions by the dashed arrows with the respective colors (Figure \ref{fig:S147}). The two kick velocities are of about equal magnitude, as scaled in equation (\ref{eq:V1jet}). The kick velocities are combined to the final NS kick velocity (red-solid arrow on Figure \ref{fig:S147}). We expect several other pairs of much weaker jets to have contributed to the explosion process. 

The pairs of jets imparting the NS kick velocity in SNR S147 are powerful. We rescaled the gravitational wave properties that \citet{Soker2023GW} estimated to these powerful pairs of jets. We draw the results by the horseshoe-shaped yellow zone on Figure \ref{Fig:StrainFrequency}. The turbulence inside the jet-inflated bubbles emits the gravitational waves \citep{Gottliebetal2023}. Such powerful jets that might impart large kick velocities lead to gravitational-wave signals that current detectors can identify from CCSNe in the Galaxy and the Magellanic Clouds (Section \ref{sec:GW}).

In the kick-BEAP mechanism, the kick is along the axis of the pair of jets, which is the momentary angular momentum of the accreted mass onto the NS. The pair of jets might compress two opposite zones, as we suggest is the case with the two zones on the ends of the cyan lines on Figure \ref{fig:S147}, or inflate ears (bubbles), as the two ears at the ends of the yellow line on Figure \ref{fig:S147}. We attribute the shaping of the pair of ears to two pairs of jets, two of several that exploded the star in the frame of the JJEM.    
We note that, in principle, the ears could have been formed by pre-explosion mass loss, and the ejecta could have expanded into the CSM that already had two ears, as suggested for some type Ia SNRs (e.g., \citealt{TsebrenkoSoker2015, Soker2022RAADelay, Soker2024RAAG19}). We note that in the two Ia SNRs with prominent ears, Kepler SNR and SNR G1.9+0.3, the ears are approximately equal in size and do not exhibit any signatures within the main SNR. In most CCSNRs with pairs of large ears, the two opposite ears are not equal, and in some cases, they have signatures that extend deep into the main shell, e.g., SNR G309.2-00.6 (for an image, e.g., \citealt{Gaensleretal1998}) and the Vela SNR (for discussion see \citealt{SokerShishkin2025Vela}, and for an image \citealt{Mayeretal2023}). 
In most CCSNe, the axis of the CSM is not aligned with the axis of the ejecta (e.g., discussion by  \citealt{Vasylyevetal2025}), as is most clearly evident in SN 1987A.

We analyzed the CCSNR morphologies by visual inspection and classification, a powerful method for studying AGN (e.g., \citealt{Hortonetal2025}) and planetary nebulae (e.g., \citealt{Sahaietal2011}). Our study further strengthens this method as a powerful tool for analyzing SNRs, providing insight into their explosion mechanism. In this study, we strengthened the kick-BEAP mechanism in the frame of the JJEM.

\begin{acknowledgments}
\added{We thank an anonymous referee for very helpful suggestions that improved the presentation of our results. }

We thank Mr. Christian Koll for allowing us to use his image of S147.

The eROSITA data shown here were processed using the eSASS software system developed by the German eROSITA consortium. \\
This work is partially based on data from eROSITA, the soft X-ray instrument aboard SRG, a joint Russian-German science mission supported by the Russian Space Agency (Roskosmos), in the interests of the Russian Academy of Sciences represented by its Space Research Institute (IKI), and the Deutsches Zentrum für Luft- und Raumfahrt (DLR). The SRG spacecraft was built by Lavochkin Association (NPOL) and its subcontractors, and is operated by NPOL with support from the Max Planck Institute for Extraterrestrial Physics (MPE). The development and construction of the eROSITA X-ray instrument was led by MPE, with contributions from the Dr. Karl Remeis Observatory Bamberg \& ECAP (FAU Erlangen-Nuernberg), the University of Hamburg Observatory, the Leibniz Institute for Astrophysics Potsdam (AIP), and the Institute for Astronomy and Astrophysics of the University of Tübingen, with the support of DLR and the Max Planck Society. The Argelander Institute for Astronomy of the University of Bonn and the Ludwig Maximilians Universität Munich also participated in the science preparation for eROSITA.
\end{acknowledgments}

\facilities{eROSITA, 
            ING:Newton}

\software{Astropy \citep{astropy:2013, astropy:2018, astropy:2022},
          eSASS4-DR1 \citep{Brunner_eSASS_2022},
          OpenCV \citep{opencv_library},
          Matplotlib \citep{Hunter:2007},
          Point source removal \citep{ShishkinSoker2025G321}
          }

\bibliographystyle{aasjournal}

\appendix
\setcounter{figure}{0}      

\renewcommand{\thesection}{\Alph{section}}

\section{Appendix: Age estimate}
\label{appendix:Appendix_Age}

\renewcommand\thefigure{A.\arabic{figure}}

\begin{figure*}
\begin{center}
\includegraphics[trim=0cm 1.5cm 0cm 2.5cm, clip, width=\textwidth]{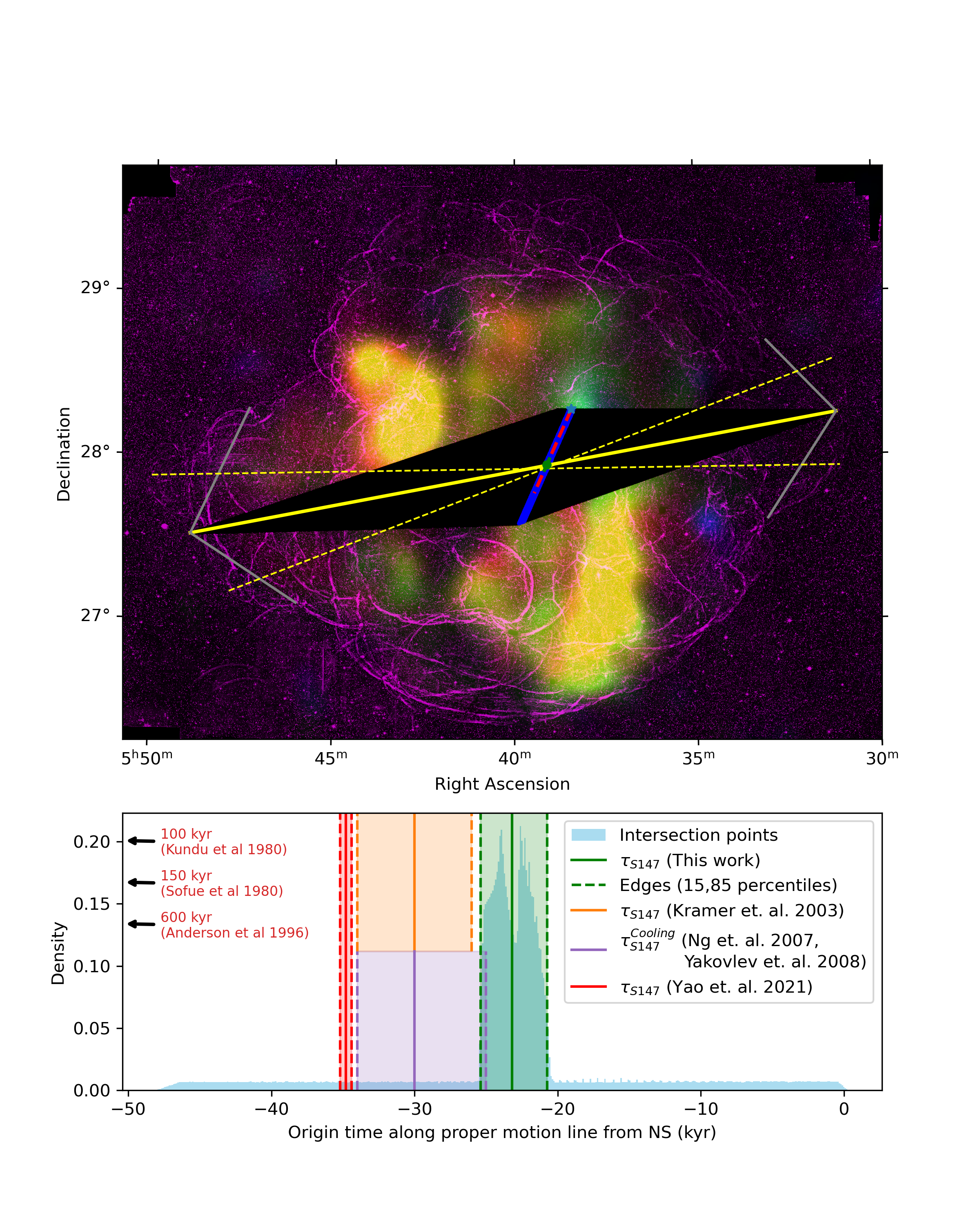} 
\caption{Our prescription for assessing the age of the remnant from the NS trajectory and the main symmetry axis. \textbf{Upper panel:} Combined X-ray and H$\alpha$ image of S147 (as Fig~\ref{fig:S147}d). As in Figures~\ref{fig:S147morphology},\ref{fig:S147}, we mark the main symmetry axis connecting the eastern and western ears with a solid-yellow line, and the two sub-ears with dashed-yellow lines. We outline the approximate extents of the two sub-ear structures with solid-gray lines. Black dots (composing a black tetragon) denote intersection points between lines connecting the sub-ears, originating along the extend of the smaller ears, and blue dots denote the intersection points that fall in the vicinity (within $50\ \rm arcsec$) of the NS trajectory (on the plain of the sky), marked with a dashed-red line. \textbf{Lower panel:} A histogram of the intersection points from the upper panel (light blue), along the NS trajectory in $\rm kyr$, as derived by the proper motion estimate of \cite{Yao_etal_2021_3Dspinkick}. We take the median and 15th and 85th percentiles as our estimate and confidence intervals, respectively. We mark the other age estimates of \cite{Kramer_etal_2003_center} in orange, of \cite{Ng_etal_2007_NSpm} and \cite{Yakovlev_etal_2008_NScool} combined in purple, and of \cite{Yao_etal_2021_3Dspinkick} in red. The vertical axis for the other age estimates serves only for visual comparison. Other age estimates from the remnant expansion rate (\citealt{Kundu_etal_1980_S147}, \citealt{Sofue_etal_1980_S147}) and NS spin-down rate (\citealt{Anderson_etal_1996_S147}) are longer, and are marked as arrows on the left.}
\label{fig:AppendixAge}
\end{center}
\end{figure*}

Using the main symmetry axis (Sec~\ref{subsec:MorphologyS147}) of SNR S147 and the trajectory of the NS at its center (J0538-2817; \citealt{Chatterjee_etal_2009_NSpm}), we examine the possible intersection points to estimate the location of the explosion and from that and the proper motion the system age. In Figure~\ref{fig:AppendixAge} we display a visual representation of the intersection points analysis (top panel) and a comparison between our result and previous estimates (lower panel).
As there is uncertainty in the orientation of the line connecting the two ears, which may differ from what we drew using the dashed-yellow lines, we consider other lines connecting the two ears. Namely, we consider possible symmetry axes extending either from the north half of the western ear to the south half of the eastern ear, or from the south half of the western ear to the north half of the eastern ear (Bottom-left to Top-right and Bottom-right to Top-left in the figure). 
We assign equal weight to all possible end points, and denote these with solid gray lines in the upper panel.
We then consider all possible intersections between pairs of lines, one from bottom-left to top-right and the other from bottom-right to top-left. We drew $10^6$ pairs of lines (axes) and their intersections form the black tetragon. We draw the dashed-yellow lines from Figure~\ref{fig:S147morphology} to illustrate one such pair of lines that gives one intersection point.
From the collection of intersection points, we examine those that fall within $50\ \rm arcsec$ of the NS trajectory (dashed-red line) as possible explosion locations (blue dots that form the blue line).

We draw the histogram of these triple-intersecting points, of the two symmetry axes, and the proper motion direction, in the lower panel of Figure~\ref{fig:AppendixAge}. The median is selected as the proposed age (solid-green line), and the 15th and 85th percentiles (corresponding to the edges of the Tophat-like distribution) as our confidence interval (dashed-green lines). The vertical axis is the intersection points density (`Freedman–Diaconis' binned). The horizontal axis is transformed from pixel distance of the points from the NS current location along the trajectory to a proposed age using the proper motion velocity. 
We find the age to be $\tau_{S147}=23.2^{+2.2}_{-2.5}\ \rm kyr$, and denote this age along the NS trajectory in the upper panel with a green line and dot. In the lower panel, we also display other age estimates for the explosion; these vary in method, but can be categorized into three classes:
\begin{itemize}
    \item \textbf{Geometrical:}
    \begin{itemize}
        \item \cite{Kramer_etal_2003_center}: $\tau_{S147}^{Kramer\ 2003}=30\pm4\ \rm{kyr}$
        \item \cite{Yao_etal_2021_3Dspinkick}: $\tau_{S147}^{Yao\ 2021}=34.8\pm0.4\ \rm{kyr}$
        \item Our work: $\tau_{S147}^{Shishkin\ 2025}=23.2^{+2.2}_{-2.5}\ \rm kyr$
    \end{itemize}
    \item \textbf{Expansion\footnote{As these are parameterized to depend on the SNR radius, we've scaled the original published estimates to use the recent radius calculation of $R_s=32.1\pm4.8\ \rm pc$ \citep{Yao_etal_2021_3Dspinkick}.}:}
    \begin{itemize}
        \item \cite{Kundu_etal_1980_S147}: $\tau_{S147}^{Kundu\ 1980}\approx100\ \rm kyr$
        \item \cite{Sofue_etal_1980_S147}: $\tau_{S147}^{Sofue\ 1980}\approx150\ \rm kyr$
    \end{itemize}
    \item \textbf{NS properties:}
    \begin{itemize}
        \item Spin-down; \cite{Anderson_etal_1996_S147}: $\tau_{S147}^{Anderson\ 1996}\approx600\ \rm kyr$ 
        \item Cooling; \cite{Ng_etal_2007_NSpm, Yakovlev_etal_2008_NScool}: $\tau_{S147}^{Cooling}=30_{-5}^{+4}\ \rm kyr$ \footnote{We've expanded the range of \cite{Yakovlev_etal_2008_NScool} to include the $\approx25\ \rm kyr$ estimate of \cite{Ng_etal_2007_NSpm} as the latter also give a higher, more recent, estimate for the NS temperature.} 
    \end{itemize}
\end{itemize}
Our age estimate is similar to other `geometrical' estimates, but it removes the dependence on a single point (geometric center) and instead explores multiple possible symmetry axes in comparison to the absolute measured quantity of the NS's proper motion. We note a possible caveat to this method here. The NS's proper motion is extrapolated for the local frame of motion, to which the SNR (i.e., the original star) might also be in motion. We estimate this to be no more than $\rm few\times10\ km/s$ (e.g., \citealt{Anguiano_etal_2020_VDF} for a general estimate, and \citealt{Dinceletal2015} for the local S147 environment), hence negligible compared to the NS projected velocity of $391\pm56 \rm \ km/s$ \citep{Yao_etal_2021_3Dspinkick}, and well within our confidence interval. For other remnants where the NS proper velocity is slower, we would have to take this into account.

\end{document}